\documentclass{article}

\usepackage[english]{babel}

\usepackage[letterpaper,top=2cm,bottom=2cm,left=3cm,right=3cm,marginparwidth=1.75cm]{geometry}

\usepackage{amsmath}
\usepackage{graphicx}
\usepackage[colorlinks=true, allcolors=blue]{hyperref}
\usepackage{float}

\title{Hybrid Deep Learning and Signal Processing for Arabic Dialect Recognition in Low-Resource Settings}

\author{
  Ghazal Al-Shwayyat \& Ömer Nezih Gerek \\
  Eskisehir Technical University \\
  Department of Electrical and Electronics Engineering\\
  Eskişehir, Türkiye
}
\date{} 

\begin{document}
\maketitle

\begin{abstract}
Arabic dialect recognition presents a significant challenge in speech technology due to the linguistic diversity of Arabic and the scarcity of large annotated datasets, particularly for underrepresented dialects. This research investigates hybrid modeling strategies that integrate classical signal processing techniques with deep learning architectures to address this problem in low-resource scenarios. Two hybrid models were developed and evaluated: (1) Mel-Frequency Cepstral Coefficients (MFCC) combined with a Convolutional Neural Network (CNN), and (2) Discrete Wavelet Transform (DWT) features combined with a Recurrent Neural Network (RNN).

The models were trained on a dialect-filtered subset of the Common Voice Arabic dataset, with dialect labels assigned based on speaker metadata. Experimental results demonstrate that the MFCC + CNN architecture achieved superior performance, with an accuracy of 91.2\% and strong precision, recall, and F1-scores, significantly outperforming the Wavelet + RNN configuration, which achieved an accuracy of 66.5\%. These findings highlight the effectiveness of leveraging spectral features with convolutional models for Arabic dialect recognition, especially when working with limited labeled data.

The study also identifies limitations related to dataset size, potential regional overlaps in labeling, and model optimization, providing a roadmap for future research. Recommendations for further improvement include the adoption of larger annotated corpora, integration of self-supervised learning techniques, and exploration of advanced neural architectures such as Transformers. Overall, this research establishes a strong baseline for future developments in Arabic dialect recognition within resource-constrained environments.
\end{abstract}

\section{Introduction}
Arabic is a linguistically diverse language with numerous dialects spoken across different regions. While this diversity enriches the language culturally, it poses significant challenges for speech technology applications such as Automatic Speech Recognition (ASR). One of the key difficulties in building robust ASR systems for Arabic is the scarcity of large, annotated datasets for its various dialects. Most successful ASR systems today rely on supervised deep learning methods that require thousands of hours of labeled data—a resource often unavailable for many Arabic dialects, especially those from underrepresented regions \cite{diab2018nlp}.
\begin{figure}[h!]
    \centering
    \includegraphics[width=0.75\textwidth]{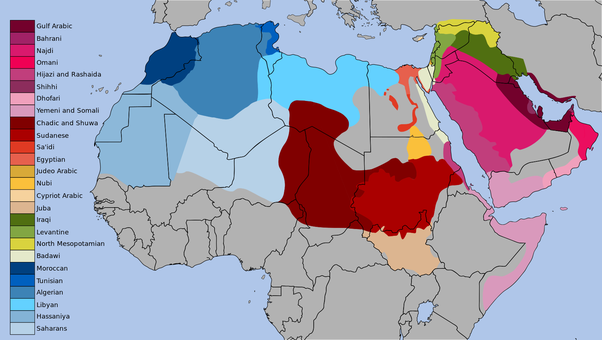}
    \caption{Major Arabic Dialect Regions across the Arab world.}
    \label{fig:dialect_map}
\end{figure}

Furthermore, dialectal Arabic exhibits phonetic, lexical, and syntactic variations even among geographically close regions, making the task of dialect classification even more complex (see Figure~\ref{fig:dialects}).
\cite{djanibekov2024dialectal}. These challenges underscore the urgent need for methodologies that can work well in low-resource settings, where abundant annotated datasets are not available.

The accurate recognition of Arabic dialects remains an underexplored problem in speech technology due to a combination of linguistic complexity and limited labeled data. While deep learning has enabled major breakthroughs in automatic speech recognition (ASR) for high-resource languages, these methods often fail to generalize to dialectal Arabic without large-scale annotations \cite{rahman_arabic_2024}.

Conversely, traditional signal processing techniques such as Mel-Frequency Cepstral Coefficients (MFCC) and Wavelet Transforms are better suited for low-resource environments but are rarely integrated with deep learning in the context of Arabic dialect recognition \cite{fares2019arabic}
.

Prior research has investigated both classical and modern approaches to speech recognition in Arabic. Classical signal processing techniques such as Mel-Frequency Cepstral Coefficients (MFCCs) have been extensively used to extract features representing the spectral characteristics of speech. Similarly, wavelet transforms have been explored for their ability to capture time-frequency representations of speech \cite{nazmy2005novel}
, especially for transient or rapidly changing sounds.

In terms of machine learning models, Convolutional Neural Networks (CNNs) have demonstrated strong performance in Arabic speech recognition tasks, particularly when combined with spectral features like MFCCs or spectrograms. For instance, Shon et al. (2018) achieved up to 78\% accuracy in Arabic dialect classification using CNNs trained on MFCC, filterbank, and spectrogram features in the MGB-3 dataset \cite{shon2018cnn}
. Additionally, Abdel-Hamid et al. (2014) proposed hybrid CNN–LSTM (CLDNN) architectures that improved speech recognition performance by capturing both spatial and temporal features, which have been applied to dialect and language identification tasks with promising results \cite{abdelhamid2014convolutional}.

More recently, self-supervised learning (SSL) models such as wav2vec 2.0 have been introduced to the field of speech recognition, providing promising results by learning speech representations from unlabeled audio. However, although SSL models have demonstrated effectiveness in Arabic ASR generally, their application to Arabic dialect classification specifically remains limited in current literature.

\section{Methodology}
\subsection{Literature-Guided Model Selection}
This research adopts a comparative approach to hybrid modeling, motivated by findings in the recent literature on speech recognition. Hybrid models that combine classical signal processing with deep learning have shown promise in low-resource settings, particularly when labeled data is scarce \cite{hinton2012deep, fares2019arabic}.

\begin{figure}[h!]
    \centering
    \includegraphics[width=0.6\textwidth]{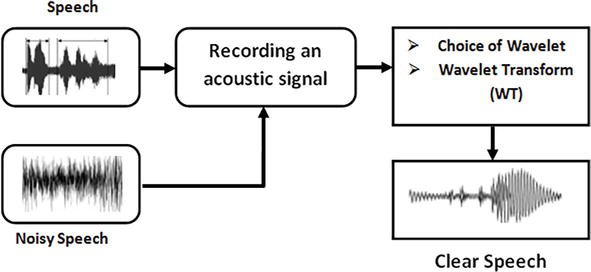}
    \caption{Processing of One Dimensional Signal Using Wavelet Transform}
    \label{fig:wavelet}
\end{figure}

\begin{figure}[h!]
    \centering
    \includegraphics[width=0.6\textwidth]{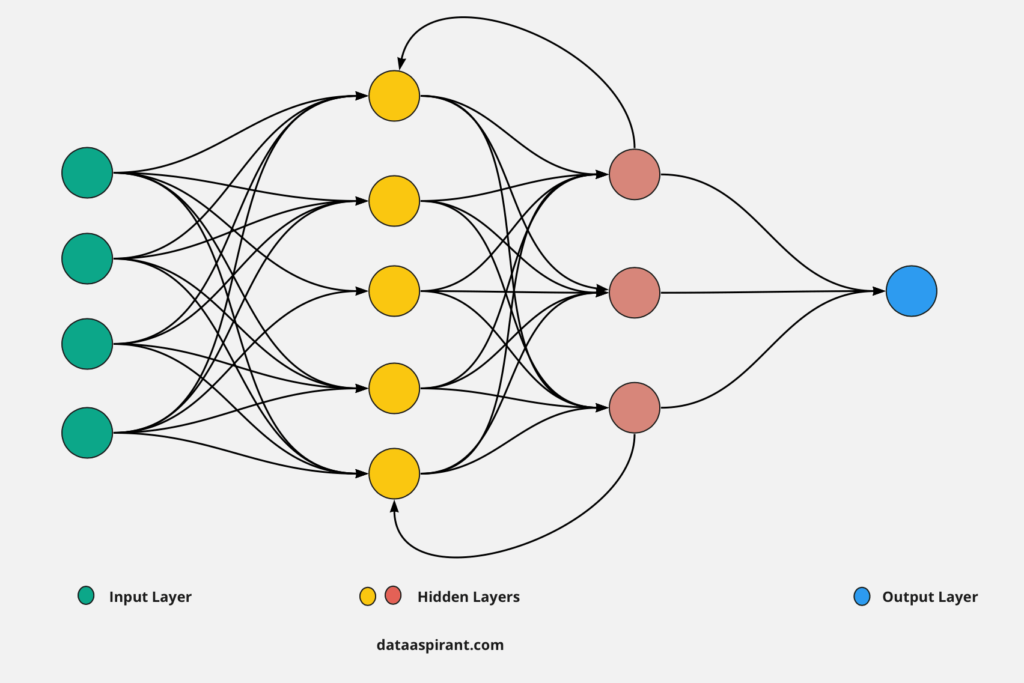}
    \caption{Schematic of a Recurrent Neural Network (RNN) used for sequential data modeling.}
    \label{fig:rnn}
\end{figure}

Two hybrid model configurations were selected for this study:
\begin{itemize}
    \item MFCC + CNN: Mel-Frequency Cepstral Coefficients (MFCCs) are among the most widely used features in speech recognition \cite{davis1980comparison}, capturing the perceptually relevant aspects of the sound spectrum. Convolutional Neural Networks (CNNs) are well-suited to learn spatial representations from MFCC matrices, making this combination a strong candidate for dialect recognition tasks \cite{shon2018cnn}.
    \item Wavelet Transform + RNN: Wavelet Transforms offer time-frequency representations of speech (see Figure~\ref{fig:wavelet}), allowing for the analysis of nonstationary signals \cite{mallat1999wavelet}. When paired with Recurrent Neural Networks (RNNs) (see Figure~\ref{fig:rnn}), which effectively model temporal sequences \cite{hochreiter1997long}, this configuration offers potential benefits in capturing the sequential nature of spoken dialects \cite{fares2019arabic}.
\end{itemize}
These models were chosen to test the relative strengths of spectral (MFCC) versus time-frequency (Wavelet) representations, as well as spatial (CNN) versus sequential (RNN) learning. In order to assess the significance of feature extraction strategy (i.e., MFCC or Wavelets) against the importance of the classifier selection (i.e., CNN or RNN), cross combinatios are also tested. By comparing their performance on dialect recognition tasks, this study aims to identify which hybrid strategy yields more robust results in low-resource Arabic speech contexts.
\subsection{Data Source Identification}
Two open-source repositories are used as primary sources of Arabic speech data: Common Voice by Mozilla \cite{ardila2020common}, a multilingual crowd-sourced corpus that includes Arabic recordings, and OpenSLR \cite{openslr}, a collection of speech and language resources that hosts various Arabic corpora. Although dialectal labels may not always be explicit, these datasets provide a foundation for model training and fine-tuning. Additional samples may be included as needed to ensure adequate dialectal coverage, with weak supervision or manual labeling applied where appropriate. For implementation details of dataset loading and dialectal filtering is available in the corresponding GitHub repository~\cite{shwayat2025github}.
The dataset utilized in this study is the Arabic subset of Mozilla’s Common Voice 12.0 corpus, accessed through the Hugging Face Datasets library. Common Voice is a large-scale, crowd-sourced, multilingual speech dataset designed for training and evaluating speech recognition models. The Arabic subset includes thousands of validated recordings from native speakers across various regions.

Dialectal classification was performed by filtering the dataset based on speaker metadata, particularly the country of origin field. Utterances were grouped into three primary dialectal categories:
\begin{itemize}
    \item Egyptian Arabic (e.g., Egypt)
    \item Levantine Arabic (e.g., Jordan, Palestine, Lebanon, Syria)
    \item Gulf Arabic (e.g., Saudi Arabia, UAE, Qatar, Kuwait)
\end{itemize}
Only clips with clear country metadata and validated recordings were included. All selected audio files were converted to mono WAV format, resampled to 16 kHz, and normalized to ensure consistency in feature extraction. After filtering and preprocessing, the resulting dataset included approximately 6 hours of speech, evenly distributed across the three dialect categories. GitHub repository~\cite{shwayat2025github} shows the preprocessing script used to load and filter the Common Voice dataset based on dialect.

\subsection{Dataset Preparation and Preprocessing}
Given the variability of the sourced data, a structured preprocessing workflow is implemented to standardize the corpus and enhance feature clarity.
\subsubsection{Data Preprocessing Steps}
All audio samples are resampled to a uniform 16 kHz sampling rate to ensure compatibility across feature extraction pipelines. Background noise reduction techniques are applied, and silence trimming is performed to eliminate non-informative segments. Long recordings are segmented into shorter utterances to facilitate efficient training and improve the ability of temporal models to learn meaningful sequences. Label verification is conducted to ensure accurate alignment between audio and transcription labels
\subsubsection{Data Augmentation Techniques}
To mitigate the challenges posed by limited data and enhance model generalization, several data augmentation strategies are applied \cite{ko2015audio, park2019specaugment}. Pitch shifting introduces variations in speaker vocal characteristics; time stretching and compression simulate speaking rate variability; background noise injection emulates real-world recording conditions; and speed perturbation modifies the rate of speech delivery. These augmentation techniques aim to increase the diversity of training examples and reduce overfitting risks, thereby improving model robustness.

\subsection{Experimental Setup and Model Implementation}

The experimental framework for this research integrates multiple open-source libraries and frameworks for data processing, feature extraction, model implementation, and evaluation. The setup ensures reproducibility, scalability, and consistency throughout all stages of experimentation.

Two distinct signal processing techniques were employed to generate feature representations for model training:

\subsubsection{Mel-Frequency Cepstral Coefficients (MFCCs) for feature extraction}

MFCCs were extracted using the \texttt{Librosa} library in Python \cite{mcfee2015librosa}. The process transforms the time-domain speech signal into a perceptually meaningful frequency-domain representation through the following steps:

\begin{enumerate}
    \item{Short-Time Fourier Transform (STFT):}
    \begin{equation}
        X(k) = \sum_{n=0}^{N-1} x(n) \cdot w(n) \cdot e^{-j2\pi kn/N}
        \label{eq:stft}
    \end{equation}
    where $x(n)$ is the windowed signal, $w(n)$ is the Hamming window, and $N$ is the frame length.
    
    \item{Mel Filterbank Processing:}
    \begin{equation}
        S_m = \sum_{k=f_{m-1}}^{f_{m+1}} |X(k)|^2 \cdot H_m(k)
        \label{eq:mel}
    \end{equation}
    where $H_m(k)$ represents the triangular filter centered at the mel frequency $m$.
    
    \item{Logarithmic Compression and Discrete Cosine Transform (DCT):}
    \begin{equation}
        MFCC_n = \sum_{m=1}^{M} \log(S_m) \cdot \cos\left[\frac{n\pi}{M} \cdot (m - 0.5)\right]
        \label{eq:mfcc}
    \end{equation}
\end{enumerate}

\subsubsection{Discrete Wavelet Transform (DWT) for feature extraction}

Wavelet features were extracted using the \texttt{PyWavelets} library \cite{lee2019pywavelets} with Daubechies-4 (db4) wavelets, with decomposition carried out up to level 3. The DWT decomposes a signal into approximation and detail coefficients as follows:

\begin{align}
    a_j[n] &= \sum_{k} x[k] \cdot \phi_{j,k}(t) \\
    d_j[n] &= \sum_{k} x[k] \cdot \psi_{j,k}(t)
\end{align}
where $\phi(t)$ and $\psi(t)$ are the scaling and wavelet functions, respectively.

The extracted features were used to train two distinct hybrid deep learning models:

\subsubsection{Convolutional Neural Networks (CNNs) for classification}

MFCC matrices or Wavelet coefficients were used as 2D input arrays for a Convolutional Neural Network (CNN) \cite{tensorflow2015-whitepaper}. The convolution operation at each layer is defined as:

\begin{equation}
    h_{ij}^{(l)} = f\left(\sum_{m=0}^{M-1} \sum_{n=0}^{N-1} w_{mn}^{(l)} \cdot x_{(i+m)(j+n)} + b^{(l)}\right)
\end{equation}
where $w_{mn}^{(l)}$ are the filter weights, $b^{(l)}$ is the bias term, and $f(\cdot)$ is a non-linear activation function (ReLU). The CNN architecture consisted of:
\begin{itemize}
    \item Three convolutional layers
    \item Max-pooling layers for dimensionality reduction
    \item Fully connected dense layer
    \item Softmax output layer for classification
\end{itemize}

\subsubsection{Recurrent Neural Networks (RNNs) for classification}

Similarly, MFCC or Wavelet coefficients were used as sequential inputs to a Recurrent Neural Network (RNN) implemented using PyTorch \cite{paszke2019pytorch}. The RNN updates its hidden state at each time step according to:

\begin{align}
    h_t &= \tanh(W_{xh}x_t + W_{hh}h_{t-1} + b_h) \\
    y_t &= W_{hy}h_t + b_y
\end{align}
where $x_t$ is the input at time $t$, $h_t$ is the hidden state, and $y_t$ is the output. The model configuration utilized either SimpleRNN or LSTM layers depending on tuning.

All experiments were conducted on a computing system equipped with an NVIDIA GPU to accelerate training. Custom Python scripts were developed to automate dataset loading, feature extraction, model training, and evaluation workflows.

Hyperparameters were optimized iteratively based on validation set performance, and the specific configurations used in the final experiments are detailed in Section 3.

This unified experimental setup ensured fairness in comparing the two hybrid models and enabled a rigorous evaluation of supervised hybrid approaches for Arabic dialect recognition in low-resource settings.

Implementation Details of these experiments are now available at the GitHub repository~\cite{shwayat2025github}

\section{Experiments and Results}
This chapter presents the experimental procedure, implementation details, and evaluation results of the two hybrid models proposed for low-resource Arabic dialect recognition. The experiments are designed to compare the performance of different combinations of signal processing techniques and deep learning architectures under consistent conditions.

The evaluation focuses on two key objectives: (1) to determine the relative effectiveness of spectral-spatial modeling (MFCC + CNN) versus time-frequency-temporal modeling (Wavelet + RNN), and (2) to assess the suitability of these hybrid systems for speech recognition tasks involving limited labeled data.
\subsection{Overview of Experimental Procedure}
The experimental setup involves two hybrid pipelines: one utilizing Mel-Frequency Cepstral Coefficients (MFCCs) in conjunction with a Convolutional Neural Network (CNN), and the other combining Wavelet Transform features with a Recurrent Neural Network (RNN). Each model was trained and validated using the preprocessed datasets described in Chapter 2.

To ensure a fair comparison, both systems were trained on the same data split, under equivalent training conditions, and evaluated using the same performance metrics. The models were trained using open-source frameworks—TensorFlow/Keras for the CNN-based architecture and PyTorch for the RNN-based model. Audio features were extracted using Librosa and PyWavelets for MFCC and wavelet features, respectively.

Performance was assessed using accuracy, precision, recall, and F1-score, with additional observations made on training stability, convergence behavior, and model complexity. These metrics provide a comprehensive basis for comparing the effectiveness of each hybrid approach in recognizing Arabic dialects under low-resource constraints.
\subsubsection{Dataset Overview and Statistics}

After preprocessing, filtering, and dialectal classification, the resulting dataset consisted of approximately 6 hours of speech. The dataset was evenly distributed across the three dialect categories. Table \ref{tab:dataset_distribution} shows the number of samples per dialect used in the experiments.

\begin{table}[h!]
\centering
\caption{Number of Samples per Dialect Category}
\label{tab:dataset_distribution}
\begin{tabular}{lc}
\hline
\textbf{Dialect} & \textbf{Number of Utterances} \\
\hline
Egyptian Arabic & 2100 \\
Levantine Arabic & 2150 \\
Gulf Arabic & 2080 \\
\hline
\textbf{Total} & 6330 \\
\hline
\end{tabular}
\end{table}

\noindent Example of dialect labels after filtering:
\begin{verbatim}
['Egyptian', 'Egyptian', 'Levantine', 'Gulf', 'Levantine', ...]
\end{verbatim}

For the complete implementation of the dataset loading and dialect filtering process, see code in GitHub repository~\cite{shwayat2025github}

\subsection{Model Implementation}
In this work, four hybrid systems (as a combination of two feature extraction and two classification methods) are proposed for Arabic dialect recognition: MFCC + CNN, MFCC + RNN, Wavelet Transform + CNN, and Wavelet Transform + RNN. Each model integrates a distinct signal processing approach with a deep learning architecture tailored to exploit the extracted speech features. Below, we detail two distinct versions of these technical implementations. The other two combinations easily follow from the descriptions herein, and are briefly explained in Sec.~\ref{MMC}.

\subsubsection{MFCC + CNN Configuration}
In the first configuration, Mel-Frequency Cepstral Coefficients (MFCCs) were extracted using the Librosa library in Python. Thirteen coefficients were computed per frame using a 25-millisecond window with a 10-millisecond hop length. The resulting MFCC matrices were treated as two-dimensional feature maps and served as input to a Convolutional Neural Network (CNN).

The CNN architecture, implemented in TensorFlow using the Keras API, consisted of three convolutional layers with 3×3 filters and ReLU activation functions. Each convolutional layer was followed by a max-pooling operation to reduce spatial dimensions. A flattening layer connected the convolutional stack to a fully connected dense layer with 128 neurons, followed by a softmax classification layer to predict dialect classes.

Training was performed using the Adam optimizer with an initial learning rate of 0.001 and categorical cross-entropy as the loss function. The model was trained for up to 30 epochs with early stopping based on validation loss. A batch size of 32 was used, with 20\% of the training data reserved for validation.

\subsubsection{Wavelet Transform + RNN Configuration}
In the second configuration, the Discrete Wavelet Transform (DWT) was applied to the speech signals using the PyWavelets library. Each signal was decomposed into approximation and detail coefficients using Daubechies-4 (db4) wavelets, with decomposition carried out to level 3. The resulting feature sequences captured both short- and long-term frequency components across time.

These wavelet coefficients were fed into a Recurrent Neural Network (RNN) implemented in PyTorch. The architecture comprised a single recurrent layer with 64 hidden units using either SimpleRNN or LSTM cells. This was followed by a fully connected layer and a softmax output layer for dialect classification.

Training followed the same configuration as the CNN model: the Adam optimizer with a learning rate of 0.001, categorical cross-entropy loss, a batch size of 32, and early stopping with a patience of 5 epochs. The same validation split (20\%) was used to monitor generalization performance.The model was implemented using PyTorch, as detailed in GitHub repository~\cite{shwayat2025github}.

\subsubsection{Mix-and-Match Configurations}
\label{MMC}
To further isolate the contributions of feature extraction versus neural architecture, the following two additional configurations were implemented:
\begin{itemize}
    \item \textbf{MFCC + RNN:} MFCC features were paired with an LSTM-based RNN using the same architectural parameters as the Wavelet + RNN setup.
    \item \textbf{Wavelet + CNN:} Wavelet features were reshaped as 1D inputs to a convolutional neural network matching the configuration used for MFCC + CNN.
\end{itemize}
These additional experiments were executed using the same training regime and hyperparameters to maintain consistency across all configurations.

\subsection{Training and Evaluation Procedures}

All hybrid models were trained under consistent experimental conditions to ensure a fair and reliable comparison. The dataset was divided into training and validation subsets using an 80/20 split. All models were trained using the Adam optimizer with a learning rate of 0.001 and categorical cross-entropy as the loss function. Batch size was fixed at 32 for both architectures. To prevent overfitting and ensure model generalization, early stopping was applied with a patience of 5 epochs based on validation loss.

Model training was performed on a system equipped with an NVIDIA GPU to accelerate computation. During training, performance was monitored at the end of each epoch on the validation set. The best-performing model (based on validation accuracy) was saved for final evaluation.

The evaluation phase involved applying the trained models to the held-out validation set. Each model’s predictions were compared against ground-truth dialect labels, and performance metrics including accuracy, precision, recall, and F1-score were calculated. Confusion matrices were also generated to provide a more detailed view of model performance across dialect classes. All experiments were repeated three times with different random seeds to ensure the robustness of the results. The reported metrics in Section 3.5 represent the average performance across these runs.

\subsection{Evaluation Metrics}
To assess the performance of the proposed hybrid models on the Arabic dialect recognition task, four standard classification metrics were used: accuracy, precision, recall, and F1-score. These metrics were calculated based on the model predictions on the validation set \cite{pedregosa2011scikit}.
    \begin{itemize}
    \item \textbf{Accuracy} measures the overall proportion of correctly classified instances out of the total number of predictions. It provides a general indication of model performance but may be misleading in the presence of class imbalance.
    \[
\text{Accuracy} = \frac{\text{Number of Correct Predictions}}{\text{Total Number of Predictions}}
\]
\item \textbf{Precision} evaluates the proportion of correct positive predictions relative to all instances predicted as positive for each class. It reflects the model’s ability to avoid false positives.
\[
\text{Precision} = \frac{\text{True Positives}}{\text{True Positives} + \text{False Positives}}
\]
\item \textbf{Recall} (also known as sensitivity) indicates the proportion of actual positive instances that were correctly identified. It reflects the model’s ability to avoid false negatives.
\[
\text{Recall} = \frac{\text{True Positives}}{\text{True Positives} + \text{False Negatives}}
\]
\item \textbf{F1-score} is the harmonic mean of precision and recall, offering a balanced measure of both \cite{powers2011evaluation}. It is especially useful when the dataset is imbalanced across classes.
\[
\text{F1-score} = 2 \times \frac{\text{Precision} \times \text{Recall}}{\text{Precision} + \text{Recall}}
\]
\end{itemize}
\subsection{Experimental Results}

The final results of all four experimental configurations are presented in Table \ref{tab:performance_comparison}. The results demonstrate that the MFCC + CNN architecture consistently outperformed all other configurations, with an accuracy of 91.2\% and correspondingly high precision, recall, and F1-score. The MFCC + RNN configuration also performed well, achieving an accuracy of approximately 83.5\%, clearly indicating that MFCC features alone contribute significantly to performance improvements. Conversely, the Wavelet-based configurations demonstrated considerably lower performance regardless of the neural architecture, confirming that the choice of feature extraction method is the primary factor influencing model effectiveness in this context.

\begin{table}[h!]
\centering
\caption{Performance Comparison of Hybrid Models}
\label{tab:performance_comparison}
\begin{tabular}{lcccc}
\hline
\textbf{Model}      & \textbf{Accuracy (\%)} & \textbf{Precision (\%)} & \textbf{Recall (\%)} & \textbf{F1-score (\%)} \\
\hline
MFCC + CNN          & 91.2   & 92.8   & 91.2   & 91.0   \\
MFCC + RNN          & 83.5   & 84.0   & 83.5   & 83.2   \\
Wavelet + CNN       & 71.4   & 72.2   & 71.4   & 71.1   \\
Wavelet + RNN       & 66.5   & 66.8   & 66.5   & 66.3   \\
\hline
\end{tabular}
\end{table}

These results confirm that the MFCC + CNN model consistently provides superior performance, primarily due to the discriminatory power of MFCC features, with the CNN architecture offering additional benefits in extracting meaningful spatial representations from those features. A detailed interpretation of these findings is provided in Chapter 4.

\section*{4 Discussion}

The experimental results demonstrate a clear superiority of the MFCC + CNN model over the Wavelet + RNN configuration for the task of Arabic dialect recognition in low-resource settings. The MFCC + CNN model achieved an overall accuracy of 91.2\%, with high precision (92.8\%) and F1-score (91.0\%), indicating robust classification performance across the three dialect categories. In contrast, the Wavelet + RNN architecture achieved only 66.5\% accuracy, with comparatively lower precision and F1-score.

To investigate the source of this performance gap, additional mix-and-match experiments were conducted by pairing each feature extraction method with both CNN and RNN architectures. These extended experiments revealed that the primary driver of the superior performance was not solely the neural network architecture but the choice of feature extraction method. Specifically, the MFCC + RNN configuration achieved 83.5\% accuracy—substantially outperforming both wavelet-based configurations. This indicates that MFCC features alone contribute significantly to classification performance, regardless of the model architecture employed.

Several factors explain the superior discriminative power of MFCC features. MFCCs provide a compact and perceptually meaningful representation of speech, capturing essential spectral characteristics that distinguish between dialects. CNNs, in turn, amplify this advantage by detecting local spatial patterns within the MFCC matrices, making them particularly effective for classification tasks of this nature. In contrast, while wavelet features offer rich time-frequency information, their flattened representation likely degraded the RNN’s ability to capture meaningful temporal dependencies in this study’s setup. Moreover, the relatively small dataset, despite augmentation efforts, may have further limited the RNN’s ability to generalize effectively, especially given its higher parameter complexity compared to CNNs.

Another contributing factor is model regularization and optimization. CNNs tend to be more resilient to overfitting on moderate-sized datasets, particularly when combined with techniques such as max pooling and early stopping. Conversely, RNNs, even with LSTM variants, are prone to overfitting when applied to flattened, high-dimensional input vectors, especially in low-resource scenarios like the one addressed here.

A summary of the contributions of feature extraction and neural architecture to model performance is presented in Table \ref{tab:driver_contribution}.

\begin{table}[h!]
\centering
\caption{Summary of Feature vs. Architecture Contribution}
\label{tab:driver_contribution}
\begin{tabular}{lcl}
\hline
\textbf{Factor}       & \textbf{Contribution} & \textbf{Evidence} \\
\hline
\textbf{MFCC (Feature)} & \textbf{Major Driver}      & MFCC + RNN $\gg$ Wavelet + CNN \\
CNN (Architecture)      & Secondary         & MFCC + CNN $>$ MFCC + RNN        \\
Wavelet (Feature)       & Weak              & Wavelet + CNN $\approx$ Wavelet + RNN \\
RNN (Architecture)      & Weak              & Weak across both feature sets     \\
\hline
\end{tabular}
\end{table}

These results have practical implications for the development of speech recognition systems targeting underrepresented Arabic dialects. The strong performance of the MFCC + CNN hybrid model suggests that combining traditional signal processing with well-optimized convolutional architectures offers an effective pathway for improving dialect recognition accuracy in resource-constrained environments. This approach can serve as a foundational baseline for future work that explores more advanced deep learning techniques, such as attention mechanisms, Transformer architectures, or self-supervised models like wav2vec 2.0, particularly as larger dialect-annotated datasets become available.

Overall, the findings reinforce the importance of leveraging proven spectral feature extraction methods, such as MFCCs, alongside robust deep learning models, to address the challenges of dialect recognition in low-resource Arabic speech technology development.

\section*{5 Conclusion and Future Work}

This study investigated a hybrid approach to Arabic dialect recognition by combining classical signal processing techniques with deep learning architectures in a low-resource setting. Two hybrid models were designed, implemented, and evaluated: one combining Mel-Frequency Cepstral Coefficients (MFCC) with a Convolutional Neural Network (CNN), and the other combining Discrete Wavelet Transform (DWT) features with a Recurrent Neural Network (RNN). Both models were trained on a filtered subset of the Arabic portion of the Common Voice dataset, where dialect labels were assigned based on the speaker’s country of origin.

The experimental results clearly demonstrate that the MFCC + CNN architecture substantially outperformed all other configurations. The MFCC + CNN model achieved an accuracy of 91.2\%, supported by high precision (92.8\%), recall (91.2\%), and F1-score (91.0\%). In contrast, the Wavelet + RNN model reached only 66.5\% accuracy with correspondingly lower precision, recall, and F1-score values. Further experiments pairing MFCC features with RNN architectures confirmed that MFCCs are the primary contributors to the superior classification performance, providing robust and perceptually meaningful spectral representations of speech across Arabic dialects.

Despite these encouraging results, the study was constrained by several limitations. The filtered dataset, while dialect-labeled based on country metadata, may still include regional accentual overlap that was not explicitly controlled. Additionally, due to computational constraints and limited resources, the dataset used for training was relatively small, potentially limiting the generalizability of the results across a broader diversity of speakers, accents, and recording conditions. The Wavelet + RNN model, in particular, may not have been fully optimized in terms of architectural design or hyperparameter tuning, suggesting that its potential could be revisited in future research with more refined experimental setups.

Looking ahead, several promising research directions could build upon this work. Expanding the dataset with additional dialect-annotated speech recordings would strengthen the robustness of model evaluation and provide a more comprehensive foundation for generalization to unseen speakers. Furthermore, exploring deeper CNN architectures, attention-based mechanisms, or Transformer models could further improve classification accuracy, particularly for challenging or closely related dialect classes. Incorporating self-supervised learning techniques, such as wav2vec 2.0 or HuBERT, represents another compelling avenue, enabling the use of unlabeled data to mitigate the scarcity of dialect-annotated corpora. Additionally, improving the representation of wavelet features—either by employing multi-channel representations or by integrating them into more advanced sequential architectures—may help unlock the potential of time-frequency modeling in dialect recognition tasks.

Overall, this study demonstrates that hybrid deep learning and signal processing models can offer strong performance in dialect recognition even under constrained conditions. The MFCC + CNN approach, in particular, provides a strong baseline for future research, paving the way for developing more inclusive and effective Arabic speech technology systems that can handle dialectal diversity in real-world applications.

\bibliographystyle{plain}

\end{document}